\documentclass{PHYEAUTH}
\usepackage{graphicx}
\usepackage{amsmath}
\usepackage{amssymb}

\begin{document}

\begin{frontmatter}

\title{
Quantum Hall Effect in a Two-Dimensional Electron System Bent by 90$^\circ$}

\author[address1]{M.~Grayson\thanksref{thank1}},
\author[address1]{D.~Schuh},
\author[address1]{M.~Bichler},
\author[address1]{G.~Abstreiter},
\author[address2]{L.~Hoeppel},
\author[address2]{J.~Smet},
and
\author[address2]{K.~von~Klitzing} 

\address[address1]{Walter Schottky Institut, Technische Universitaet Muenchen, D-85748 
Garching, Germany}

\address[address2]{Max-Planck-Institute fuer Festkoerperforschung, D-70569 Stuttgart, 
Germany}

\thanks[thank1]{
Corresponding author. 
E-mail: mgrayson @alumni.princeton.edu
Fax: +49 89 320 6620}

\begin{abstract}

Using a new MBE growth technique, we fabricate a two-dimensional electron
system which is bent around an atomically sharp 90$^\circ$ corner.  In the
quantum Hall regime under tilted magnetic fields, we can measure
equilibration between both co- and counter-propagating edge channels of
arbitrary filling factor ratio.  We present here 4-point magnetotransport
characterization of the corner junction with filling factor combinations
which can all be explained using the standard Landauer-B\"uttiker edge
channel picture.  The success of this description confirms the realization
of the first non-planar quantum Hall edge geometry.

\end{abstract}

\begin{keyword}

% keywords here, in the form: keyword \sep keyword
Quantum Hall effect \sep edge states \sep co-propagating \sep 
counter-propagating \sep corner quantum well

% PACS codes here, in the form: \PACS code \sep code
\PACS 72.20.My \sep 72.15.Gd \sep 73.40.Lq \sep 73.40.Hm

\end{keyword}
\end{frontmatter}

%[main text]
\section{Introduction}

In the quantized Hall regime, a clean two-dimensional electron system
(2DES) with electron density $n$ in the presence of a magnetic field $B$
exhibits quantized steps in the Hall resistance near values of magnetic
field where the filling factor, $\nu = hn/eB$ is an integer or
odd-denominator fraction, describing the integer \cite{vonKlitzing} or
fractional \cite{Tsui} quantum Hall effect, respectively.  These steps
result from a mobility gap in the bulk region, leaving gapless edge modes
responsible for describing current conduction in the system
\cite{Halperin}.  Early \cite{Haug,vanWees,Chang,Kouwenhoven} and recent
\cite{Wuertz,Deviatov} experiments on the quantum Hall effect tested the
equilibration properties of edge channels in gated structures that
populate different channels at different chemical potentials.  However,
the reduced mobility of low-density gated regions restricts the filling
factor combinations that can be tested, and in planar structures co- and
counterpropagating equilibration studies require fundamentally different
gate designs \cite{Wuertz,Deviatov}.  We present in this paper a new type
of device for studying edge state equilibration which permits abrupt
junctions between arbitrary filling factors of both co- {\it and}
counter-propagating edges, simply by tilting the sample in a magnetic
field.

\begin{figure}[h]
%h=here, t=top, b=bottom, p=separate figure page
\begin{center}
\leavevmode
\includegraphics[width=0.7\linewidth]{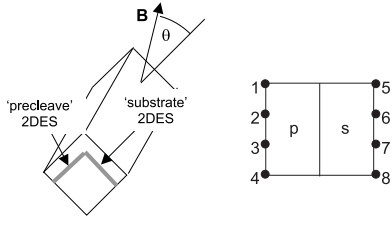}
\caption{Schematic of the overgrown corner showing the substrate
and precleave 2DES's (left) as well as a top perspective of the
contact layout (right).}
\label{figurename}
\end{center}
\end{figure}

\section{Sample}

We call our device the corner-quantum well heterojunction (CQW),
fabricated by overgrowing a standard GaAs/AlGaAs heterojunction structure
on a precleaved corner \cite{Grayson1} as depicted schematically in Fig.
1, left.  The length of the corner junction in the data presented here is
$L = $ 3.2~mm, and indium contacts are alloyed to each facet away from the
corner junction, and indexed as shown in the top perspective of Fig. 1,
right.  We identify one side of the device as the 'substrate' (s), and the
other side as the 'precleave' (p) as in Fig. 1 \cite{Grayson1}, and
measure slightly different electron densities for the two facets under
different cool-downs as listed in the captions of Fig. 2 and Fig. 3.  We
designate the contact configuration of all 4-point resistance measurements
with the notation $R_{i-j,k-l}$ for current leads $i-j$ and voltage
measured between $k-l$.

\begin{figure}[b]
%h=here, t=top, b=bottom, p=separate figure page
\begin{center}
\leavevmode
\includegraphics[width=1.0\linewidth]{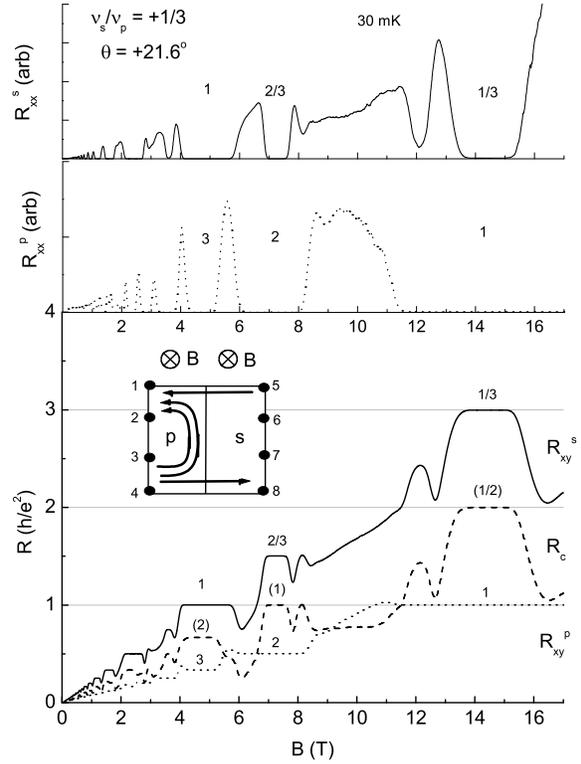}
\caption{Plot of $R_{xx}^s = R_{5-8,6-7}$ (top), $R_{xx}^p = R_{1-4,2-3}$
(middle), $R_{xy}^s = R_{2-6,7-5}$, $R_c = R_{2-6,1-5}$, and
$R_{xy}^p = R_{2-6,3-1}$ (bottom) measurements of the corner well at 
tilted $B$ fields such that $\nu_s/\nu_p = +1/3$. $n_s = 1.07 \times
10^{11} {\rm cm}^{-2}$ and $n_p = 1.30 \times 10^{11} {\rm cm}^{-2}$.  
Plateaus in $R_{xy}$ and $R_c$ are indexed with quantum number, $n$
according to $R = h/n e^2$.  The bottom figure demonstrates the
Landauer-B\"uttiker prediction $R_c = |R_{xy}^s| - |R_{xy}^p|$.}
\label{figurename}
\end{center}
\end{figure}

\begin{figure}[b]
%h=here, t=top, b=bottom, p=separate figure page
\begin{center}
\leavevmode
\includegraphics[width=1.0\linewidth]{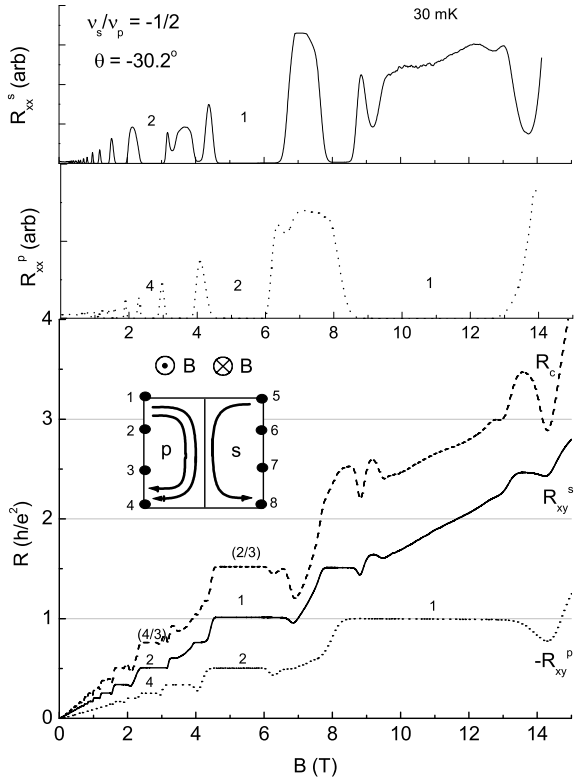}
\caption{
Plot of $R_{xx}^s = R_{5-8,6-7}$ (top), $R_{xx}^p = R_{1-4,2-3}$
(middle), $R_c = R_{2-6,1-5}$,  $R_{xy}^s = R_{2-6,7-5}$ and
$-R_{xy}^p = R_{2-6,1-3}$ (bottom) measurements of the corner well at
tilted $B$ fields such that $\nu_s/\nu_p = -1/2$. $n_s = 1.15 \times
10^{11} {\rm cm}^{-2}$ and $n_p = 1.25 \times 10^{11} {\rm cm}^{-2}$.
Plateaus in $R_{xy}$ and $R_c$ are indexed with quantum number, $n$
according to $R = h/n e^2$.  The bottom figure demonstrates the
Landauer-B\"uttiker prediction $R_c = |R_{xy}^s| + |R_{xy}^p|.$}
\label{figurename}
\end{center}
\end{figure}

\section{Results}

In the presence of a tilted magnetic field at an angle $\theta$ relative
to the substrate normal (Fig. 1, left) the relative filling factor between
the two systems $\nu_s/\nu_p$ can be tuned according to:

\begin{equation}
\frac{\nu_s}{\nu_p}=\frac{n_s/B cos(\theta)}{n_p/B sin(\theta)}
=\frac{n_s}{n_p}tan(\theta)
\end{equation}

\noindent The high quality of the growth is demonstrated by fractional
quantum Hall effect (FQHE) minima appearing on both facets below 1 K.  
For angles $0^\circ < \theta < +90^\circ ~ (\nu_s/\nu_p > 0)$ the edge
channels counter-propagate at the corner reminiscent of the coupling
chirality between gated QHE regions in standard planar structures.  But
for $0^\circ > \theta > - 90^\circ ~ (\nu_s/\nu_p < 0)$, the normal
component of the magnetic field {\it changes sign} across the junction,
resulting in a junction of {\it co-propagating} edge states of arbitrary
filling factor simply by tilting the junction in a B-field.

We begin our Hall measurements in the $\nu_s/\nu_p > 0 $ regime.  Fig. 2,
top, shows $R_{xx}^p = R_{1-4,2-3}$ and $R_{xx}^s = R_{5-8,6-7}$ for
$\nu_s/\nu_p = +1/3$, corresponding to a magnetic field tilt angle of
$\theta = +21.6^\circ$.  The minimae of the series $(\nu_s:\nu_p) =$
(1:3), $(\frac{2}{3}:2)$, $(\frac{1}{3}:1)$ go to zero for both $R_{xx}$
traces, validating the use of the edge channel picture at these fields.  
The inset of Fig. 2 shows a cartoon of the (1:3) case, showing the sign of
the normal $B$-field component in each region.  In Fig. 2, bottom,
$R_{xy}^s = R_{2-6,7-5}$ and $R_{xy}^p = R_{2-6,3-1}$ are plotted along
with the 4-point corner resistance $R_c = R_{2-6,1-5}$.  Because current
contact number 2 feeds both voltage contacts 4 and 8, the 4-point
resistance $R_c' = R_{2-6,4-8} = 0$ is trivially zero.  By inspection
then, $R_c$ is just the difference between the two Hall resistances, $R_c
= |R_{xy}^s| - |R_{xy}^p|$ confirming the Landauer-B\"uttiker picture
\cite{Buettiker} with full equilibration of edge channels on reaching
contact 1.

Next we examine $\nu_s/\nu_p < 0 $, the condition novel to this paper
where the normal magnetic field changes sign across the junction, and the
edge channel in the precleave system correspondingly switches chirality
(Fig. 3, inset).  Fig. 3, top, shows $R_{xx}^p = R_{1-4,2-3}$ and
$R_{xx}^s = R_{5-8,6-7}$ for $\nu_s/\nu_p = -1/2$, corresponding to a
$B$-field tilt angle of $\theta = -30.2^\circ$.  There we see
well-developed minimae at $(\nu_s:\nu_p) =$ (2:4) and (1:2).  In this
novel geometry, the co-propagating edge channels at the $L = 3.2$ mm long
corner fully equilibrate, meaning the downstream potential at contacts 4
and 8 are equal and $R_c' = R_{2-6,4-8} = 0$.  Correspondingly, the only
non-trivial 4-point resistance is again $R_c = R_{2-6,1-5}$.  With the
opposite chirality in the two systems, we note that since $R_{xy}^p$ has
changed polarity $R_c$ is now the {\it sum} of the individual $R_{xy}$'s:  
$R_c = |R_{xy}^s| + |R_{xy}^p|$.  It is worth noting that in planar gate
defined geometries, no 4-point resistance can be {\it larger} than the
largest individual $R_{xy}$, so we have successfully demonstrated a new
regime of validity for the Landauer-B\"uttiker formalism.

It remains to be seen what physics will be observed at junction lengths $L
\sim l_{eq}$ of order the interchannel equilibration length, where
outgoing channels from the corner are not fully equilibrated, but rather
partially reflected/transmitted.  In such a case, dependences on co-
versus counter-propagation as well as on the specific value $(\nu_s :
\nu_p)$ are anticipated, and at a lossless junction, novel effects like a
dc step-up transformer are predicted, due to Andreev-like quasiparticle
reflection \cite{Safi,Chklovskii,Sandler}.  Future experiments with small
junctions can also measure interchannel equilibration lengths for both co-
and counterpropagating edges in the same device.

\section{Conclusion}

In conclusion, we observe quantum Hall effect in a new corner quantum well
structure.  The relative filling factor between 2DEGs on the two facets of
the corner is tunable by tilting the magnetic field angle, and two
different regimes are studied where the edge channels at the corner
junction are co- and counter-propagating.  This paper determines that the
device functions as expected, and is well behaved in the quantum Hall
regime.

\ack{This work is supported financially by the Deutsche Forschungs
Gemeinschaft via Schwerpunktprogramm "Quanten-Hall-Systeme".}

\end{document}